\documentclass[11pt]{article}
\usepackage[margin=1in]{geometry}

\usepackage{amsmath,amssymb}
\usepackage{authblk}
\usepackage{tikz-cd}
\usepackage{multicol}
\usepackage{cite}

\newcommand{\Tr}{\mathrm{Tr}}

\title{\textbf{Proof of Quantum Conformal Invariance in $\mathcal N=4$ Super-Yang-Mills Theory via $\mathcal N=1$ Superfields}}
\author{Akihisa D.-E. Tateishi}
\date{}
\affil{\itshape RIKEN, 2-1 Hirosawa, Wako, Saitama 351-0198, Japan}

\begin{document}
\maketitle

\begin{abstract}
Using the $\mathcal{N}=1$ superfield formalism, we prove that the superconformal symmetry of $\mathcal{N}=4$ super-Yang-Mills theory is preserved in the quantum theory. We demonstrate that the $\mathcal{N}=1$ calculation is sufficient to guarantee the full $\mathcal{N}=4$ supersymmetry, and show that the results are exact without receiving any higher-order corrections.
\end{abstract}
\newpage

\section{Introduction}
Supersymmetric field theories have long been a central focus of theoretical physics. In particular, four-dimensional $\mathcal N=4$ super-Yang-Mills (SYM) theory possesses many intriguing properties, such as S-duality (Montonen-Olive duality)~\cite{Montonen:1977sn, Vafa:1994tf}, which extends electromagnetic and strong-weak dualities; the AdS/CFT correspondence~\cite{Maldacena:1997re, Witten:1998qj}, which closely relates $\mathcal N=4$ SYM to type IIB supergravity on $AdS_5 \times S^5$; integrability in the planar limit~\cite{Minahan:2002ve, Beisert:2010jr}; and the double copy structure~\cite{Bern:2008qj, Arkani-Hamed:2013jha}, which relates the amplitudes of $\mathcal{N}=8$ supergravity to the square of those in $\mathcal N=4$ SYM. In this paper, we investigate the exact superconformal invariance of the theory at the quantum level.

On the other hand, explicit evaluation of the path integral is a fundamental approach to understanding quantum field theories. For example, the chiral anomaly in gauge theories is calculated by explicitly performing the path integral~\cite{Fujikawa:1979ay}, as are its $\mathcal N = 1$~\cite{Konishi:1985tu, Konishi:1983hf} and $\mathcal{N} = 2$~\cite{Tateishi:2025ay} supersymmetric extensions. In addition, supersymmetric localization~\cite{Witten:1988ze, Nekrasov:2002qd, Pestun:2007rz, Kapustin:2009kz}, which reduces supersymmetric infinite-dimensional path integrals to finite-dimensional integrations over the moduli space, has emerged as a powerful tool for exact computations.

In this paper, we prove the exact quantum superconformal invariance of $\mathcal{N}=4$ super-Yang-Mills theory. While this result has previously been confirmed using component fields~\cite{Jones:1977am, Poggio:1977tu, Avdeev:1982jx}, our approach utilizes the $\mathcal{N}=1$ superfield formalism to evaluate the path integral. This formulation allows us to manifest the background supersymmetry throughout the calculation, demonstrating that the superconformal anomaly vanishes directly. Furthermore, we demonstrate how this $\mathcal{N}=1$ framework is sufficient to guarantee the full $\mathcal{N}=4$ supersymmetry, and establish that our results are exact without receiving any higher-order corrections.

This paper is organized as follows. We begin in Sec.~\ref{section:SYM_Lagrangian_transformation} by reviewing the Lagrangian and superconformal transformations of $\mathcal{N}=4$ super-Yang-Mills theory. Sec.~\ref{section:measure_variation} is devoted to the computation of the measure variation; while the analyses of individual fields in Subsecs.~\ref{subsection:variation_chiral} and \ref{subsection:variation_vector} follow established literature, the total contribution evaluated in Subsec.~\ref{subsection:variation_total} represents the original core of this work. We then discuss the preservation of the full $\mathcal{N}=4$ supersymmetry and the one-loop exactness in Sec.~\ref{section:discussion}. Lastly, we conclude with a summary in Sec.~\ref{section:summary}.

\section{$\mathcal N=4$ Super-Yang-Mills Theory\label{section:SYM_Lagrangian_transformation}}
The action of $\mathcal{N}=4$ SYM~\cite{Gates:1983nr, Tachikawa:2013kta}, expressed in terms of $\mathcal{N}=1$ superfields~\cite{Wess:1992cp}, is given by
\begin{align}
    S=\Tr\left(\frac14\int d^6z\,\mathcal EW^\alpha W_\alpha+\frac14\int d^6\bar z\,\mathcal E^\dagger\bar W_{\dot\alpha}\bar W^{\dot\alpha}+\sum_{i=1}^3\int d^8z\,E\,e^{-V}\Phi_i^\dagger e^V\Phi_i\right.\nonumber\\
    \left.+\int d^6z\,\mathcal E\Phi_1[\Phi_2,\Phi_3]+\int d^6\bar z\,\mathcal E^\dagger\Phi_3^\dagger[\Phi_2^\dagger,\Phi_1^\dagger]\right),\label{eq:N=4_SYM_action}
\end{align}
where the dynamical superfields $V$ and $\Phi_i$ are a real vector superfield and chiral superfields, respectively:
\begin{align}
  V=V^\dagger,\quad\bar D_{\dot\alpha}\Phi_i=D^\alpha\Phi_i^\dagger=0,
\end{align}
$W_\alpha$ and $\bar W_{\dot\alpha}$ are the chiral and antichiral field strengths defined by
\begin{align}
W^\alpha\equiv-\frac14\bar D^2\left(e^{-V}D^\alpha e^V\right),\qquad\bar W_{\dot\alpha}\equiv+\frac14D^2\left(e^V\bar D_{\dot\alpha}e^{-V}\right),
\end{align}
and $\mathcal{E}$ ($\mathcal{E}^\dagger, E$) is the chiral (antichiral, real) superspace density that ensures the invariance of the integration measures $d^6z\,\mathcal{E}$, $d^6\bar{z}\,\mathcal{E}^\dagger$, and $d^8z\,E$ under general coordinate transformations.
\par
At the classical level, this action enjoys the superconformal symmetry under the transformations
\begin{gather}
  \Phi_i\to e^\Sigma\Phi_i,\quad V\to V,\\
  \left(\mathcal E\to\mathcal E\,e^{-3\Sigma},\quad E\to E\,e^{-\Sigma-\Sigma^\dagger},\quad W^\alpha W_\alpha\to e^{3\Sigma}W^\alpha W_\alpha\right).
\end{gather}
In this article, we show that this invariance indeed holds true even at the quantum level.

\section{Variation of the Path Integral Measure under Superconformal Transformations\label{section:measure_variation}}
Since the classical action is invariant under the superconformal transformations, proving the symmetry at the quantum level reduces to showing the invariance of the functional measure. In this section, we compute the measure's variation to first order in the transformation parameter $\Sigma$. Expressed in terms of $\mathcal{N}=1$ superfields, the Jacobian factor is given by
\begin{align}
  J = \exp \left( - \int d^4x d^2\theta \, \Sigma \, \mathcal{A} + \text{h.c.} \right).
\end{align}
In what follows, we calculate the contributions to $\mathcal{A}$ from the vector superfield $V$ and the three chiral fields $\Phi_i$ separately.

\subsection{Chiral Superfields\label{subsection:variation_chiral}}
First, the contributions from the chiral superfields $\Phi_i$ are identical in form to the Konishi anomaly~\cite{Konishi:1985tu, Konishi:1983hf}:
\begin{align}
  \mathcal{A}_{\Phi_i} = -\frac{1}{16\pi^2} T_{\text{Adj}} \text{Tr}(W^\alpha W_\alpha),\label{eq:measure_variation_chiral}
\end{align}
where $T_{\text{Adj}}$ is the Dynkin index for the adjoint representation of the gauge group, normalized to $T_{\text{Adj}}=N$ for $\mathrm{SU}(N)$.

\subsection{Vector Superfield\label{subsection:variation_vector}}
Next, we consider the contribution from the vector superfield. Just as in the case of non-\allowbreak supersymmetric gauge theories, the vector superfield possesses gauge degrees of freedom that cause the path integral to diverge. Hence, a prescription to eliminate these degrees of freedom by introducing ghosts is required. In the background field formalism\cite{Gates:1983nr, Arkani-Hamed:1997ut, Grisaru:1982zh}, the gauge-fixed path integral measure is given by
\begin{align}
\mathcal{D}\mu = \mathcal{D}v \mathcal{D}c \mathcal{D}c' \mathcal{D}\bar{c} \mathcal{D}\bar{c}' \mathcal{D}b \mathcal{D}\bar{b}.\label{eq:functional_measure_gauged_vector_superfield}
\end{align}
Here, the quantum fluctuation $v$ is defined via the relation
\begin{align}
e^V &= e^{\Omega^\dagger} e^v e^\Omega,
\end{align}
where $V$ denotes the full (quantum) vector superfield, and the prepotential $\Omega$ is related to the background field $V_{\text{bg}}$ through
\begin{align}
e^{V_{\text{bg}}} &= e^{\Omega^\dagger} e^\Omega.
\end{align}
In the expression~(\ref{eq:functional_measure_gauged_vector_superfield}), the Faddeev-Popov ghosts $c,c',\bar c, \bar c'$ are introduced analogously to non-supersymmetric gauge theories, while the Nielsen-Kallosh ghosts $b,\bar b$ are employed to eliminate the extra Jacobian $\det(\bar{D}^2 D^2)$ arising from the path integration over the Faddeev-Popov ghosts. 

Note that the superconformal weight is $1$ for the ghosts $c,c',\bar c, \bar c', b, \bar b$, matching that of standard chiral superfields. In contrast, the weight of $v$ is determined to be $1/2$ by requiring the invariance of its quadratic action,
\begin{align}
\int d^8z \, E \, v \square_V v,
\end{align}
where $\square_V$ is the background-covariant d'Alembertian. Using these weights, we evaluate the individual contributions to the anomaly.

First, because the Faddeev-Popov ghosts $c$ and $c'$ are fermionic (anticommuting) chiral superfields, their contributions are equal in magnitude but opposite in sign to that of a dynamical chiral superfield:
\begin{align}
\mathcal A_{c}=\mathcal A_{c'}=-\mathcal A_{\Phi_i}.
\end{align}

In contrast, since the Nielsen-Kallosh ghost $b$ obeys bosonic statistics, its contribution is exactly equal to that of the dynamical field:
\begin{align}
\mathcal A_{b}=\mathcal A_{\Phi_i}.
\end{align}

Lastly, for a unit superconformal weight, the measure variation of the quantum fluctuation $v$ is exactly $-4$ times that of a chiral superfield. This reflects the fact that an unconstrained real superfield has $32$ off-shell degrees of freedom, whereas a chiral superfield has only $8$. Since the actual weight of $v$ is $1/2$, the final contribution is given by
\begin{align}
\mathcal A_v=\frac12\cdot(-4)\mathcal A_{\Phi_i}=-2\mathcal A_{\Phi_i}.
\end{align}

Combining these results, the total contribution from the gauge-fixed vector sector is given by
\begin{align}
  \mathcal A_V&=\mathcal A_v+\mathcal A_c+\mathcal A_{c'}+\mathcal A_b\\
  &=(-2-1-1+1)\mathcal A_{\Phi_i}\\
  &=-3\mathcal A_{\Phi_i}.\label{eq:measure_variation_vector}
\end{align}

\subsection{Total Contribution\label{subsection:variation_total}}
Combining the results from Eqs.~(\ref{eq:measure_variation_chiral}) and (\ref{eq:measure_variation_vector}), and taking into account the three chiral superfields in $\mathcal{N}=4$ super-Yang-Mills theory, the total variation of the path integral measure under superconformal transformations vanishes identically:
\begin{align}
\mathcal A=3\mathcal A_{\Phi_i}+\mathcal A_V=0.
\end{align}
Consequently, the superconformal symmetry of $\mathcal{N}=4$ SYM is preserved in the quantum theory, at least at the one-loop order.

\section{Discussion\label{section:discussion}}
Two remarks are in order concerning our results. First, one might wonder whether the full $\mathcal{N}=4$ supersymmetry is preserved throughout the preceding calculation. By construction, our formalism manifests only the $\mathcal{N}=1$ subgroup. However, since $\mathcal{N}=4$ super-Yang-Mills theory possesses the $\mathrm{SU}(4)_R$ $R$-symmetry that transforms the four fermions into one another, preserving the $\mathcal{N}=1$ supersymmetry is sufficient to guarantee the invariance under the full $\mathcal{N}=4$ supersymmetry.

The second remark concerns one-loop exactness. While conformal anomalies generally receive higher-order perturbative corrections, Eq.~(\ref{eq:measure_variation_chiral}) takes the exact same form as the Konishi anomaly in $\mathcal{N}=1$ gauge theories. Because the latter is non-perturbatively exact at one loop, our results are likewise protected against higher-order corrections.

Taking these two points together, we conclude that the superconformal symmetry of $\mathcal{N}=4$ SYM is preserved in the quantum theory to all orders.

\section{Summary\label{section:summary}}
In this work, we have established the exact quantum superconformal invariance of $\mathcal{N}=4$ super-Yang-Mills theory utilizing the $\mathcal{N}=1$ superfield formalism. This approach allowed us to carry out the entire analysis while keeping the background $\mathcal{N}=1$ supersymmetry manifest. Furthermore, we have shown that this $\mathcal{N}=1$ framework is sufficient to guarantee the full $\mathcal{N}=4$ supersymmetry, and that the resulting anomaly coefficients are protected against any higher-order corrections.

Beyond the scope of this work, there remain numerous intriguing avenues concerning quantum anomalies and exact path-integral computations in supersymmetric theories. It would be highly worthwhile to reformulate these non-perturbative phenomena within the superfield formalism. For instance, a promising direction is the formulation of supersymmetric localization~\cite{Pestun:2007rz}—particularly for path integrals around instanton backgrounds~\cite{Nekrasov:2002qd}—in terms of superfields, which could provide a more direct, algebraic approach to extracting instanton-corrected effective actions or moduli space metrics. Furthermore, extending our $\mathcal{N}=1$ framework to analyze multi-loop anomaly structures~\cite{Shifman:1986zi} and non-perturbative dualities~\cite{Intriligator:1995au} would be a significant step toward a deeper understanding of quantum superconformal field theories. We hope to return to these fascinating problems in future work.

\bibliographystyle{unsrt}
\bibliography{report5_N=4_SYM_conformal_invariance_references}
\nocite{Grisaru:1980nk}

\end{document}